# Breather molecular complexes in a passively mode-locked fibre laser

*Junsong Peng[1,2], Zihan Zhao[1], Sonia Boscolo[3*], Christophe Finot[4], Srikanth Sugavanam[3], Dmitry V. Churkin[5], Heping Zeng[1,2*]*

*Corresponding Author: s.a.boscolo@aston.ac.uk; hpzeng@phy.ecnu.edu.cn

[1]State Key Laboratory of Precision Spectroscopy, East China Normal University, Shanghai 200062, China

[2]Chongqing Institute of East China Normal University, Chongqing 401120, China

[3]Aston Institute of Photonic Technologies, Aston University, Aston Triangle, Birmingham B4 7ET, UK

[4]Laboratoire Interdisciplinaire Carnot de Bourgogne, UMR 6303 CNRS – Université de Bourgogne Franche-Comté, F-21078 Dijon Cedex, France

[5]Novosibirsk State University, Pirogova str. 2, Novosibirsk 630090, Russia

Abstract: Breathing solitons are nonlinear waves in which the energy concentrates in a localized and oscillatory fashion. Similarly to stationary solitons, breathers in dissipative systems can form stable bound states displaying molecule-like dynamics, which are frequently called breather molecules. So far, the experimental observation of optical breather molecules and the real-time detection of their dynamics have been limited to diatomic molecules, that is, bound states of only two breathers. In this work, we report on the observation of different types of breather complexes in a mode-locked fibre laser: multi-breather molecules, and molecular complexes originating from the binding of two breather-pair molecules or a breather pair molecule and a single breather. The inter-molecular temporal separation of the molecular complexes attains several hundreds of picoseconds, which is more than an order of magnitude larger than that of their stationary soliton counterparts and is a signature of long-range interactions. Numerical simulations of the laser model support our experimental findings. Moreover, non-equilibrium dynamics of breathing solitons are also observed, including breather collisions and annihilation. Our work opens the possibility of studying the dynamics of many-body systems in which breathers are the elementary constituents.

**1. Introduction**

The concept of soliton is universal and can be applied to a large class of solitary wave propagation phenomena that are observed in most branches of nonlinear science, ranging from fluid dynamics and biology to plasma physics and photonics [1]. The main feature of solitons is that they propagate for a long time without visible changes. In their original studies solitons were attributed to integrable Hamiltonian systems like the focusing one-dimensional nonlinear Schrödinger equation (NLSE) [2], but lately the initial concept has been extended to nonlinear dissipative systems, in which localized wave packets arise from the balance between dispersion or diffraction and nonlinearity (as the conventional solitons) and between gain and losses [3]. Besides their formation, their mutual interactions such as collisions and even the emergence of stable bound states are currently the subject of intense studies in laser physics by means of real-time ultrafast measurements [4-7]. Such dissipative multi-soliton complexes, also termed soliton molecules, result from self-organization phenomena in a laser cavity, and show dynamics similar to matter molecules, such as synthesis and vibration. In [8], through transfer of general spectroscopy concepts to the case of dissipative solitons, resonant excitation and all-optical switching of soliton molecules in response to an external perturbation were demonstrated, thus extending the matter-like soliton molecule analogy. While the soliton-pair molecule, consisting of two bound solitons, has been the most studied multi-soliton structure, a recent study showed that two soliton-pair molecules can bind subsequently to form a stable molecular complex [9].

Apart from stationary solitons, numerous nonlinear systems support breathing dissipative solitons exhibiting a periodic oscillatory behavior. Dissipative breathers were first demonstrated in passive Kerr cavities [10], and subsequently reported in optical micro-resonators [11, 12]. Dissipative systems whose averaged dynamics are governed by the complex Ginzburg-Landau equation (CGLE) also support breather solutions [13]. In the context of passively mode-locked lasers, which constitute a suitable platform for the study of the properties and dynamics of nonlinear dissipative systems, the existence of regimes of operation in which the laser oscillator generates strongly breathing solitons was recently predicted theoretically [14]. In [15], we have directly revealed the fast spectral and temporal dynamics of breathers in a mode-locked fibre laser by means of real-time detection techniques. Similarly to Akhmediev [16] and Kuznetsov-Ma [17, 18] breathers in conservative nonlinear systems, breathing dissipative solitons are related to the Fermi-Pasta-Ulam recurrence [19, 20] – a paradoxical evolution of nonlinearly coupled oscillators that periodically return to their original state. Breathing solitons are not only objects of fundamental interest in nonlinear science, but they are also attractive because of their potential for applications, such as in spectroscopy [21].

Contrary to the significant progress made in the study of the dynamics of stationary soliton molecules in ultrafast lasers, the existence of bound states of breathing solitons has been less explored. Patterns of breathers have been mostly studied in conservative systems described by the NLSE, and only breather collisions [22] and periodic breather interactions [23] have been reported. In fact, overlapping soliton pulses cannot form strong bonds in such systems as the effective interaction potential of the pulses is not a minimum [24]. A potential minimum can be induced by dissipative effects only, thus enabling the formation of multi-pulse bound states. A number of recent works have reported on the observation of breathing soliton-pair molecules in mode-locked fibre lasers [15, 25, 26]. It is worth mentioning that besides the formation of breather molecules, various notable dynamics of dissipative localized structures have been experimentally studied in mode-locked lasers, including soliton and breather explosions [27-31] and rogue wave generation [32].

In the present work, by employing an ultrafast fibre laser setup whose output is spectrally and temporally analyzed in real time, we demonstrate multi-breather molecules and two types of breather molecular complexes (BMCs). One originates from the stable binding of two basic molecules, each made up of a pair of breathing solitons ((2+2) BMC), and the other one arises from the binding of a breather-pair molecule and a single breather ((2+1) BMC). The inter-molecular temporal separation of such complexes is more than an order of magnitude larger than that of their stationary soliton counterparts. A recent theoretical study has shown that the interaction of well-separated dissipative breathers by exchanging weakly decaying dispersive waves can lead to the formation of bound states as a result of harmonic synchronization [33]. Such dispersive waves emitted by breathers in the anomalous dispersion propagation regime is the driving mechanism for the formation of the breather complexes that are demonstrated in this work. We also explore the non-equilibrium dynamics of BMCs, including collisions of breathers and annihilation of an elementary breather within a BMC. Numerical simulations of the laser model described by the CGLE confirm our experimental observations.

## 2. Results

### 2.1. Experimental setups

The experimental setup for generating breather complexes is an erbium-doped passively mode-locked fibre laser with anomalous path-averaged cavity dispersion, sketched in Fig. 1 (see "Supporting Information" for details). Mode locking relies on the nonlinear polarization evolution (NPE) technique [34], which is realized through the inclusion of two fibre polarization controllers (PCs) and a polarization dependent isolator, and allows us to tune the nonlinear transfer function by simply rotating the loops, thereby modifying the linear cavity loss, hence the interactions among pulses. The output signal from the laser is analyzed in real time in both the time and frequency domains. We use the dispersive Fourier transform (DFT) method [35, 36] to acquire sub-nanometer resolution spectral measurements at the shot-to-shot level. This simple, yet powerful method, which is today a commonly used tool to obtain spectral dynamics on ultra-short time scales, maps the optical spectrum of the laser output onto a temporal waveform that is directly read out on a real-time oscilloscope. This is achieved by stretching the laser output pulses in a dispersive medium that cumulates a group-velocity dispersion large enough for the pulse propagation to satisfy the far-field condition, hence for the stretched waveform to represent the spectral intensity of the initial pulse waveform. The time-domain dynamics of the laser are characterized by measuring time traces of one-dimensional intensity in real time, $I(t)$, and then constructing from these traces the spatio-temporal intensity evolution $I(t, z)$ [37]. The latter reveals both the dynamics over the fast time $t$ and the slow evolution propagation coordinate $z$, which is measured, in our case, as a number of cavity round trips.

## 2.2. Experimental results

The laser works in the stable single-pulse mode-locking regime, characterized by soliton-like pulse shaping, when the pump current is set to 44.9 mA. When the pump current is increased to 46 mA, self-starting mode locking is accompanied by the generation of multiple pulses per cavity round trip. The existence of the stable single-pulse regime in a narrow range of pump powers is due to the fact that the cavity length is longer than the typical length of most mode-locked fibre lasers (less than 10 m). In our experiment, the pump current is fixed at 46 mA and we tune the cavity loss through small rotations of the PCs to generate various kinds of breather complexes. We note that the laser can also sustain regimes of emission of two, three and four bound stationary solitons. Figure 2 shows the dynamics of a breather-quartet molecule ('tetratomic molecule'). The spatio-temporal intensity evolution depicted in Fig. 2a reveals large periodic variations in the intensities of the four elementary breathers with a period of approximately 1000 round trips. The peak intensities within each period of pulsation change by nearly an order of magnitude (Fig. 2b). Concomitantly, the pulse temporal duration also varies, whilst the temporal resolution of our detection system (around 30 ps) does not allow us to capture such changes. The use of a time lens [38] may potentially relax this constraint. However, this tecnhique may prove difficult to implement when the pulses experience large breathing. In spite of the periodic intensity variations, the pulse temporal separation remains nearly fixed at approximately 50 ps over consecutive cavity round trips, indicating a strong bond between the pulses. The fast evolutionary behavior of the pulse temporal intensity entails rapid variations of the optical spectrum, which are beyond the speed of traditional spectral measurement tools such as an optical spectrum analyzer. We employ real-time DFT spectral monitoring to capture the spectral evolution of the pulse quartet over cavity round trips. The DFT-based single-shot spectral measurements are shown in Fig. 2c. The spectrum features the typical interference pattern that is present in the spectrum of a soliton molecule [4], and the separation between the peaks of spectral intensity matches the pulse spacing in time. The evolution of the spectrum over cavity round trips is periodic, and the spectrum largely widens and narrows within each period, with the broadening (compression) naturally occurring in the vicinity of the position where the pulses reach the highest (lowest) peak intensity. The maximum width of the spectrum exceeds the minimum width by more than eight times (Fig. 2d). The energy of the pulse quartet (Fig. 2c, white curve), calculated by integration of the power spectral density over the whole wavelength band, evolves over round trips synchronously with the spectrum width and pulse temporal intensity. The calibration and accuracy of the DFT spectral measurements are checked by comparing, in the Supporting Information, the average of the measured consecutive single-shot spectra with the averaged spectrum recorded by an optical spectrum analyzer. To complete the quantitative dynamical picture of the breather-quartet molecule, we resolve in real-time the relative phases within the molecule using the DFT-based spectral interferometry method, which relies on computing the Fourier transform of each single-shot DFT spectrum and is now rather commonly used with stationary soliton states [9], [39]. It is worth to mention that the application of this method to the breather complexes generated in our laser cavity is quite challenging as the first-order single-shot autocorrelation traces are too weak when the breathers are at their intensity minima. The internal phase dynamics shown in the Supporting Information confirm the strong nature of the bond between the breather constituents of the molecule.

A notable phenomenon in the breathing process illustrated in Fig. 2 is the periodic appearance of wide Kelly sidebands in the shot-to-shot spectra (see, e.g., Fig. 2d). These Kelly sidebands are a manifestation of resonant dispersive waves in the temporal domain, which, in the usual perspective, radiate from solitons when they are perturbed by lumped nonlinear losses and various intra-cavity components in a round trip [40]. The low-intensity background pattern synchronized to the breather pattern that is observable in Fig. 2a is formed out of the slowly decaying dispersive waves radiated by the breathers [41], which are the main agent of the breathers long-range interaction [33]. These dispersive waves are periodically emitted during hundreds of cavity periods and, as it can be seen from Fig. 2c (a magnified version of Fig. 2c is provided in the Supporting Information) and Fig. 2d, the central wavelength of the spectrum and the wavelengths of the Kelly sidebands oscillate synchronously with the pulse energy over cavity round trips [41, 42]. This wavelength oscillation can be ascribed to cross-phase modulation of the breathers and the radiated dispersive waves [41]. As we can notice from Fig. 2a, this periodic wavelength shift is mirrored into a periodic shift of the pulse positions in the time domain. In the false color plot of Fig. 2a, the pulsations of the individual pulse intensities (round trip number from approximately 300 to 900, for example) are accompanied by a positive (to the right) time shift, while no time shift occurs over the duration of the intensity minima. As a result, the pulses at a given round trip number within a period of oscillation appear shifted in time with respect to the pulses at the same round trip number within the previous oscillation period while having the same group velocity. We note that this fact has no physical origin but, as explained in the Supporting Information, is a direct result of the spatio-temporal intensity representation. It is also noteworthy that the signal after temporal stretching has a rather long duration; for instance, the wavelength span in Fig. 2c corresponds to a time window of length 25 ns. Therefore, the impact of the artificial shift of the pulse temporal positions from one oscillation period to the next (approximately 100 ps in Fig. 2a) on the DFT measurements is negligible. The calculated physical time shift induced by the fibre dispersion over consecutive round trips shows good agreement with the artificial shift that is observed in the spatio-temporal intensity map. Although the Kerr effect can also induce a change in the pulse temporal positions, this effect is negligibly small (Supporting Information).

Another type of a breather complex is observed under different PC settings: a (2+2) BMC consisting of two bound breather pairs. The spatio-temporal intensity dynamics shown in Fig. 3a reveal the existence of two characteristic time scales, corresponding to the pulse separation within each breather-pair molecule and the separation between the two breather-pair molecules. The ratio between the two characteristic times being approximately one to seven indicates that the two times are plausibly associated with two different bond strengths. The intra-molecular pulse separation is nearly the same as that of the breather-quartet molecule in Fig. 2, but the inter-molecular separation comes to approximately 350 ps, an order of magnitude larger than that of the stationary soliton molecular complexes recently reported [9]. Through rotation of the PCs, we can turn to a different (2+2) BMC (Fig. 3b). Compared to the previous case (Fig. 3a), the pulses of the two breather-pair molecules are bound at different time intervals, showing that a BMC can have varied intra-molecular separations.

By adjusting the linear cavity loss, we can also decrease the number of breathing solitons and prepare a robust breather triplet. An example is provided in Fig. 3c, which shows the spatio-temporal evolution of a BMC consisting of a breather-pair molecule bound to a single breather, where the latter can be regarded as a monoatomic molecule. Though the single breather is far apart from the breather pair by about 500 ps, its intensity evolves synchronously with the breather pair over cavity round trips, indicating a strong inter-molecular bond. We have also observed a single breather-pair molecule in the laser (see Supporting Information).

Soliton interaction is one of the most exciting areas of research in nonlinear dynamics. In nearly conservative systems, such as Bose-Einstein condensates of atoms confined to a quasi-dimensional waveguide, soliton collisions may result in annihilation depending on the relative phase between the solitons [43]. Soliton annihilation also occurs in dissipative systems [44]. In our experiment, we have observed breather collisions and destruction within unstable BMCs by rotating the PCs at fixed pump power. Examples are shown in Fig. 4. In Fig. 4a, the initial condition is a breather quartet. Breather fusions occur at round-trip numbers of approximately 2500 and 3500, followed by the splitting of each of the two resulting breathers into two pulses at a round-trip number of approximately 5000. Subsequently, one of the two breather pairs gradually disappears. This is the first time that splitting of breathing solitons is observed, following the observation of soliton splitting [6, 45, 46]. Figure 4b shows three collision events when the initial condition is a breather triplet. In all cases, the scenario resembles that of an elastic collision, in that the two colliding breathers just pass through each other and they are not affected by the collision: they do not merge and their traces do not change after the collision. This is different from the inelastic breather collisions that were recently observed in passive fibers, which represent an example of a conservative system [22]. In that case, indeed, the two breathers merged into a single peak. It is also notable that different types of non-equilibrium dynamics of breather molecular complexes can be observed in our laser setup by fine adjustment of the PC settings. For instance, the time separation between the leading and trailing breather within a complex may increase or decrease monotonously over cavity round trips, a dynamical state that we could refer to as thermal expansion or contraction of a breather molecular complex.

**2.3. Numerical simulations**

In [41], the periodic shift of the central wavelength of the pulsating soliton spectrum as well as the appearance of wide Kelly sidebands on the spectrum were explained in the frame of a vectorial model for the laser operation, based on coupled extended NLSEs and including the effect of cross-phase modulation. Our focus is here on verifying the main features of the observed pulse complexes, that is, their periodic breathing behavior. To this end, we have bypassed a quite complex laser model and carried out numerical simulations of the laser based on the master-equation approach [47], namely, we have used the cubic-quintic CGLE. For the context of a passively mode-locked fibre laser, the cubic-quintic CGLE has the form [48]

$$i\psi_\zeta + \frac{D}{2}\psi_{\tau\tau} + |\psi|^2\psi + \nu|\psi|^4\psi = i\delta\psi + i\varepsilon|\psi|^2\psi + i\beta\psi_{\tau\tau} + i\mu|\psi|^4\psi \quad (1)$$

where $\zeta$ is the normalized propagation distance traversed in the cavity, $\tau$ is the retarded time, $\psi$ is the normalized envelope of the field, $D = -\text{sgn}(\bar{\beta}_2)$, $\bar{\beta}_2$ is the path-averaged cavity dispersion, and $\nu$ corresponds to the saturation of the nonlinear refractive index. The dissipative terms are written on the right-hand side of Eq. (1), where $\delta > 0$ ($\delta < 0$) is the net linear gain (loss) coefficient, the term with $\varepsilon$ represents the nonlinear gain (which arises, e.g., from saturable absorption), $\beta > 0$ accounts for spectral filtering, and $\mu < 0$ represents the saturation of the nonlinear gain. This equation is one of the simplest models for a passively mode-locked laser with a fast saturable absorber, and can describe various complex nonlinear dynamics such as soliton explosions [27, 28, 49], rogue waves [32], multiple pulsing [50], switching dynamics [51], and dissipative soliton resonances [52] among the others [48, 53].

By solving the CGLE numerically for the set of parameters given in the caption of Fig. 5, we have found periodic breathing dynamics of a bound pulse quartet, which qualitatively account for the behaviors observed in Fig. 2. We have also found numerically the different types of BMCs generated in the experiment (Fig. 3), as shown in Fig. 6.

## 3. Conclusion

One of the remarkable properties of dissipative solitons, which are mostly absent in integrable systems, is the ability to form robust multi-pulse bound states (soliton molecules). An optical cavity constitutes an ideal propagation medium to study multiple soliton interactions, since even very weak interactions can be revealed through the virtually unlimited propagation time [44, 54, 55]. While soliton pairs constitute the central soliton molecule case, soliton molecules can exist in various isomers [56], and a large population of optical solitons can self-assembly into macromolecules and soliton crystals [57], and even into highly-ordered supra-molecular structures through the tailoring of their long-range interactions [58]. Although breathers are fundamentally different from stationary solitons, they appear to exhibit similar collective dynamics. The experimental generation of breather-pair molecules in a fibre laser cavity has been reported on a few recent occasions [15, 25, 26]. The emphasis of the present article is on pushing the similarity in collective behavior between breathing and stationary solitons further. We have reported on the experimental observation and real-time dynamic characterization of different types of breather complexes in a passively mode-locked fibre laser, including tetratomic molecules, and molecular complexes formed by the binding of two diatomic molecules or a diatomic and a monoatomic molecule. We have also observed breather annihilation within an unstable molecular complex when the complex is in the phase of intensity drop. Since breathing dissipative solitons are fundamental modes of many nonlinear physical systems, it is reasonable to assume that the breather dynamics observed in this work will incentivize the investigation of BMCs in various other systems.

Similarly to the stationary soliton scenario [58], we believe it will be possible to assemble supra-molecular structures of optical breathers in a fibre laser by using intense opto-acoustic effects in photonic crystal fibres. So far, the studies of localized wave structures in ultrafast fibre optics are limited to the time and frequency domains, as these structures are generated in single-mode fibres. The generation and propagation of pulses in multimode fibre systems have recently drawn great attention [59]. A multimode fibre laser provides a new degree of freedom in the control of coherent light fields: the space domain. We can therefore anticipate that the spatio-temporal engineering of optical pulses will enable the generation of even more complex breather structures.

**Supporting Information**

Figure S1: Validation of the dispersive Fourier transform (DFT) implementation.

Figure S2: Characterisation of a breather-quartet molecule.

Figure S3: Evidence for the nonlinear interaction between breathing solitons and coherent dispersive waves.

Figure S4: Graphical explanation of the artefact of the spatio-temporal intensity representation.

Figure S5: Relationship between physical and artificial time shifts in the spatio-temporal intensity map.

Figure S6: Dispersion-induced time shift.

Figure S7: Dynamics of a breathing soliton-pair molecule.

**Acknowledgements**  (We acknowledge the support from the National Natural Science Fund of China (11621404, 11561121003, 11727812, 61775059, and 11704123), National Key Research and Development Program (2018YFB0407100), Key Project of Shanghai Education. Commission (2017-01-07-00-05-E00021), and Science and Technology Innovation Program of Basic Science Foundation of Shanghai (18JC1412000)).

Competing interests: There are no financial competing interests.

**Keywords**: (ultrafast fibre lasers, breathers, mode locking)

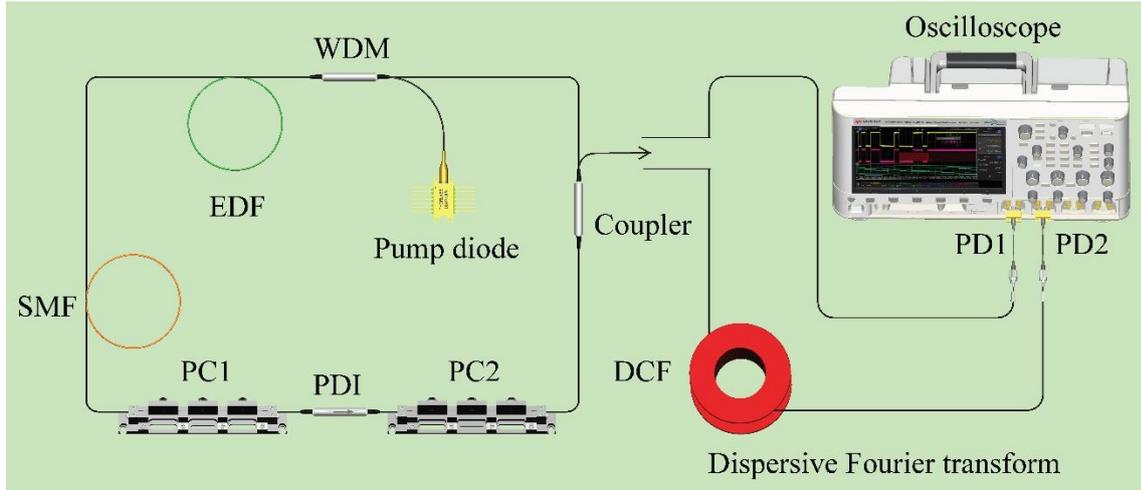

**Figure 1. Schematic of the laser and real-time detection system.** WDM, wavelength division multiplexer; PC, polarization controller; PDI, polarization dependent isolator; SMF, single mode fibre; EDF, erbium doped fibre; DCF, dispersion compensating fibre.

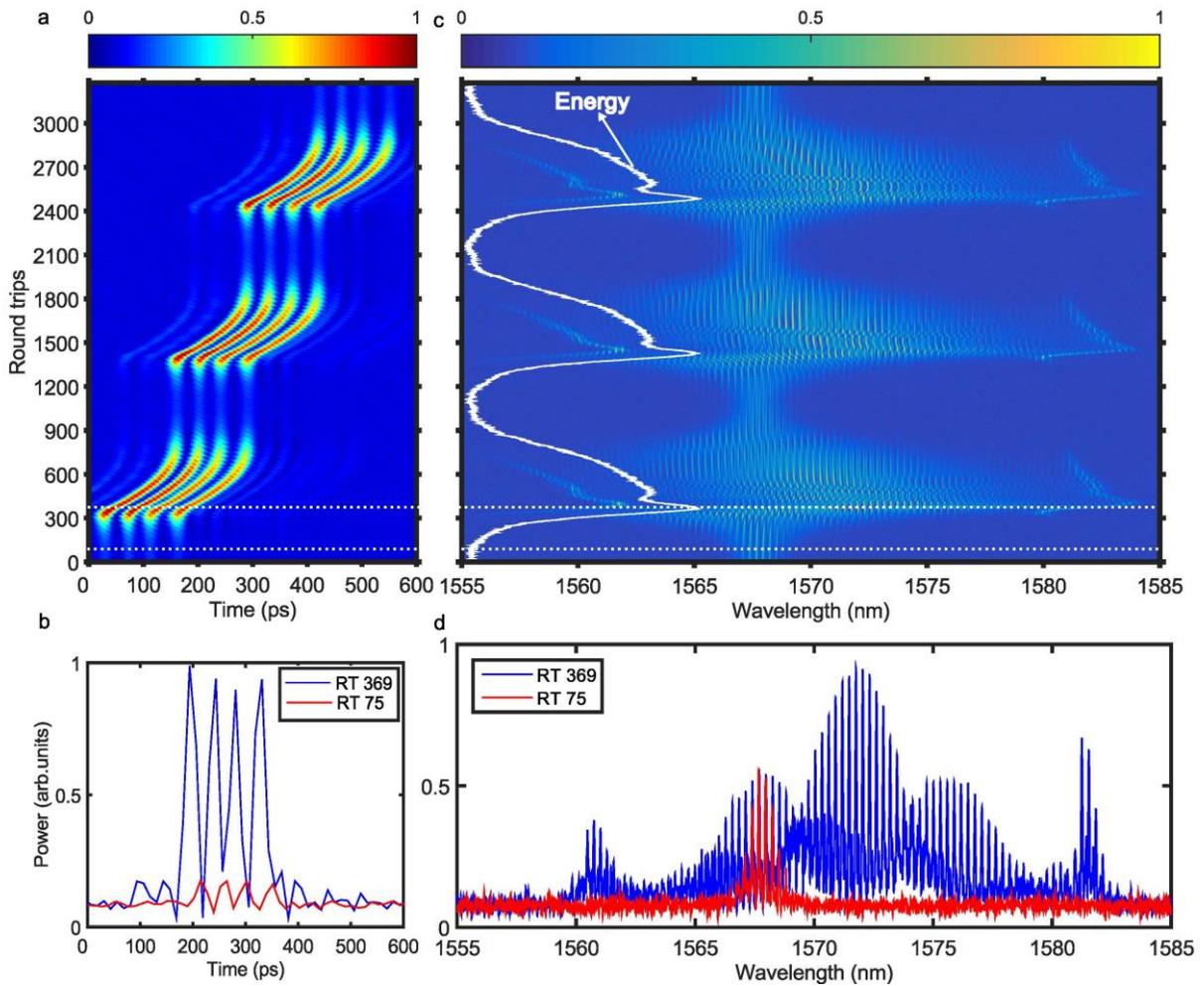

**Figure 2. Dynamics of a breathing soliton quartet molecule.** (a) Temporal evolution of the intensity relative to the average round-trip time over consecutive round trips. (b) Temporal intensity profiles at representative round-trip numbers of maximal and minimal energies within a period of oscillation. (c) DFT recording of single-shot spectra. The evolution of the energy is also shown (solid white line). (d) Single-shot spectra at representative round-trip numbers of maximal and minimal spectrum extents within a period.

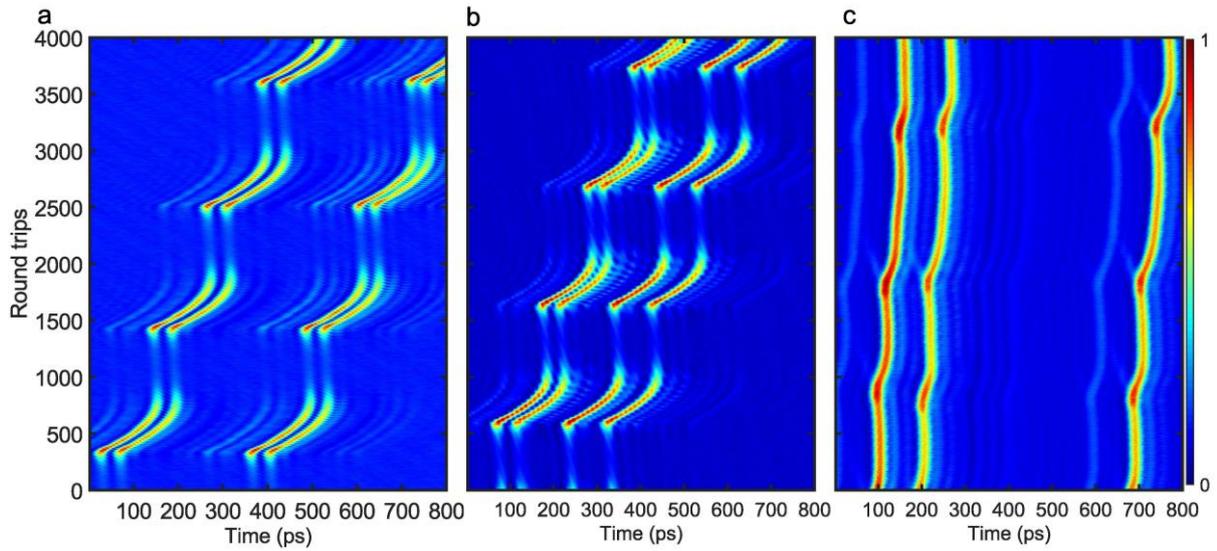

**Figure 3. Dynamics of various breather molecular complexes.** Temporal evolution of the intensity relative to the average round-trip time over 4000 consecutive round trips for: (a) a (2+2) breather molecular complex with equal intra-molecular pulse separations; (b) a (2+2) breather molecular complex with different intra-molecular separations; and (c) a (2+1) breather molecular complex.

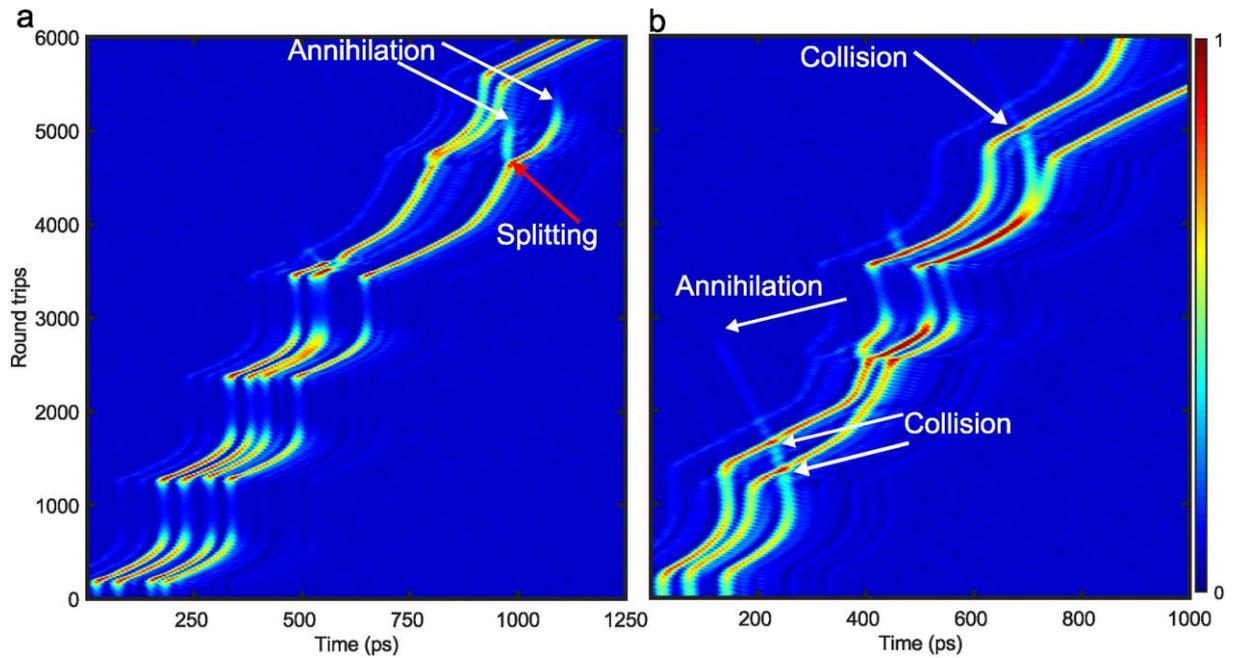

**Figure 4. Examples of non-equilibrium dynamics of breather molecular complexes**. (a) Breather splitting and annihilation of elementary breathers around a round-trip number of 5000. (b) Three collision events; the weak breather ultimately vanishes after the collisions.

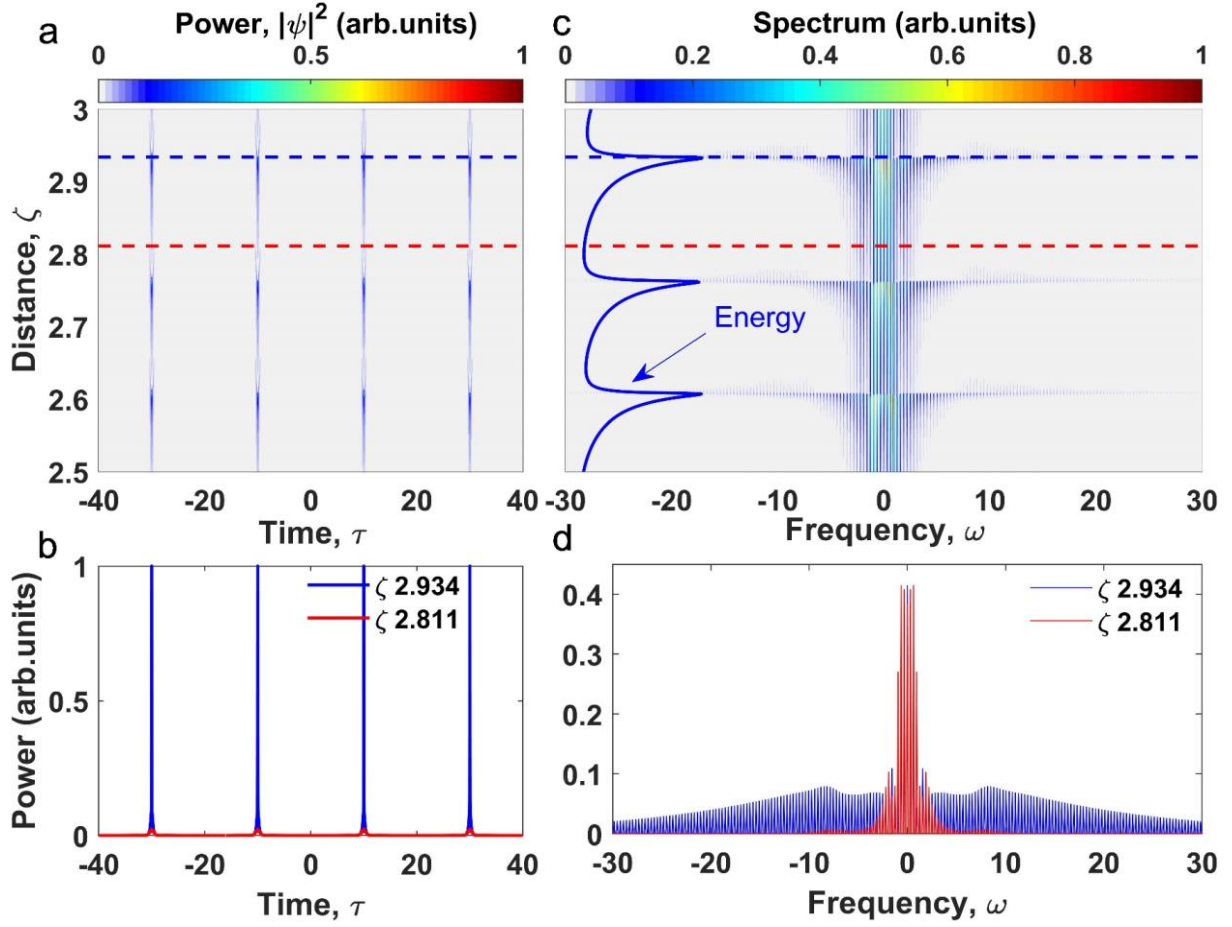

**Figure 5. Simulation of a breather-quartet molecule.** Evolutions of the (a) temporal intensity and (c) spectrum found for the set of CGLE parameters: ($\nu = 2$, $\delta = -0.01$, $\beta = 0.3$, $\varepsilon = 5$, $\mu = -0.02$). The evolution of the energy $Q(\zeta) = \int d\tau |\psi(\zeta,\tau)|^2$ is also shown. (b) Temporal intensity profiles at representative round-trip numbers of maximal and minimal energies within a period of oscillation. (d) Corresponding spectra of maximal and minimal extents.

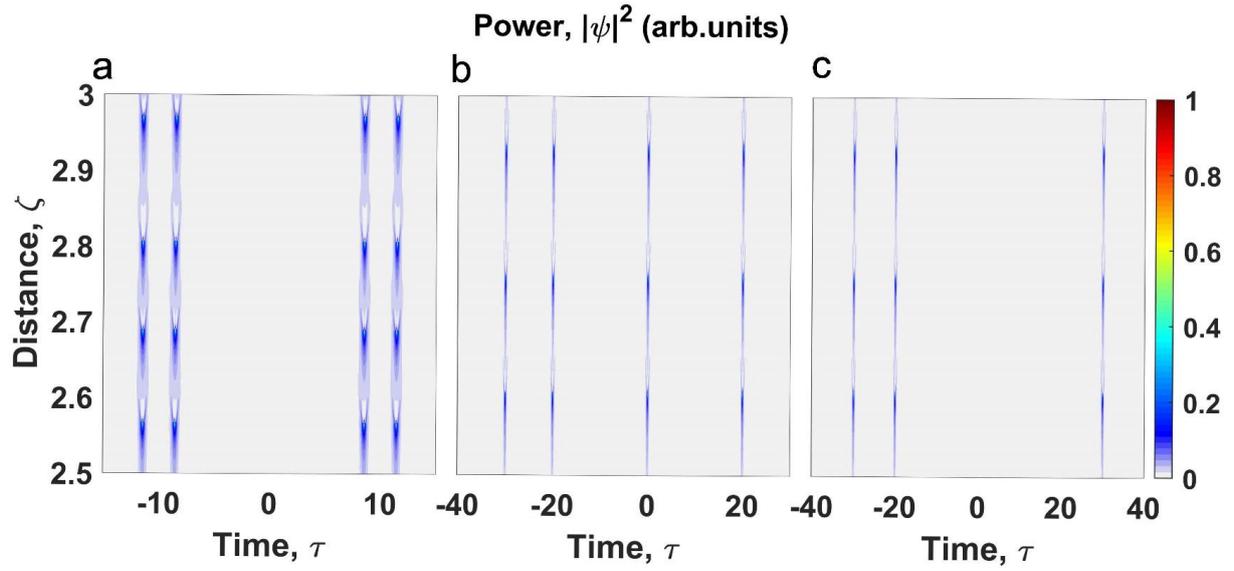

Figure 6. Simulation of various breather molecular complexes. Evolution of the temporal intensity for: (a) a (2+2) breather molecular complex with equal intra-molecular pulse separations; (b) a (2+2) breather molecular complex with different intra-molecular separations; and (c) a (2+1) breather molecular complex. The CCQGLE parameters are ($\nu = 2$, $\delta = -0.23$, $\beta = 0.4$, $\varepsilon = 6$, $\mu = -0.02$) for the solution shown in panel (a), while the same set of parameters as that of Fig. 5 is used for the solutions shown in panels (b) and (c).